\pdfoutput=1
\documentclass{article}

\usepackage{microtype}
\usepackage{graphicx}
\usepackage{subfigure}
\usepackage{booktabs} % for professional tables
\usepackage{mathrsfs}
\usepackage{hyperref}
\usepackage{amsmath,amsfonts,bm}

\def\eqref#1{equation~\ref{#1}}
\def\1{\bm{1}}

\def\vd{{\bm{d}}}

\def\vq{{\bm{q}}}

\DeclareMathAlphabet{\mathsfit}{\encodingdefault}{\sfdefault}{m}{sl}
\SetMathAlphabet{\mathsfit}{bold}{\encodingdefault}{\sfdefault}{bx}{n}

\def\emE{{E}}

\def\emM{{M}}

\def\emT{{T}}

\DeclareMathOperator*{\argmax}{arg\,max}
\DeclareMathOperator*{\argmin}{arg\,min}

\usepackage[accepted]{icml2019}

\icmltitlerunning{A Multi-Resolution Word Embedding for Document Retrieval from Large Unstructured Knowledge Bases}

\begin{document}

\twocolumn[
\icmltitle{A Multi-Resolution Word Embedding for Document Retrieval from Large Unstructured Knowledge Bases}



\icmlsetsymbol{equal}{*}

\begin{icmlauthorlist}
\icmlauthor{Tolgahan Cakaloglu}{uoac}
%\icmlauthor{Christian Szegedy}{google}
\icmlauthor{Xiaowei Xu}{uoai}
\end{icmlauthorlist}

\icmlaffiliation{uoac}{Department of Computer Science, University of Arkansas, Little Rock, Arkansas, United States}
%\icmlaffiliation{google}{Google LLC., Mountain View, California , United States}
\icmlaffiliation{uoai}{Department of Information Science, University of Arkansas, Little Rock, Arkansas, United States}

\icmlcorrespondingauthor{Tolgahan Cakaloglu}{txcakaloglu@ualr.edu}

\icmlkeywords{Machine Learning, ICML}

\vskip 0.3in
]



\printAffiliationsAndNotice{}  % leave blank if no need to mention equal contribution

\begin{abstract}
Deep language models learning a hierarchical representation proved to be a powerful tool for natural language processing, text mining and information retrieval. However, representations that perform well for retrieval must capture semantic meaning at different levels of abstraction or context-scopes. In this paper, we propose a new method to generate multi-resolution word embeddings that represent documents at multiple resolutions in terms of context-scopes. In order to investigate its performance,we use the Stanford Question Answering Dataset (SQuAD) and the Question Answering by Search And Reading (QUASAR) in an open-domain question-answering setting, where the first task is to find documents useful for answering a given question. To this end, we first compare the quality of various text-embedding methods for retrieval performance and give an extensive empirical comparison with the performance of various non-augmented base embeddings with and without multi-resolution representation. We argue that multi-resolution word embeddings are consistently superior to the original counterparts and deep residual neural models specifically trained for retrieval purposes can yield further significant gains when they are used for augmenting those embeddings.
\end{abstract}

\section{Introduction}

\begin{figure}[ht]
\vskip 0.2in
\begin{center}
\centerline{\includegraphics[width=.9999\columnwidth]{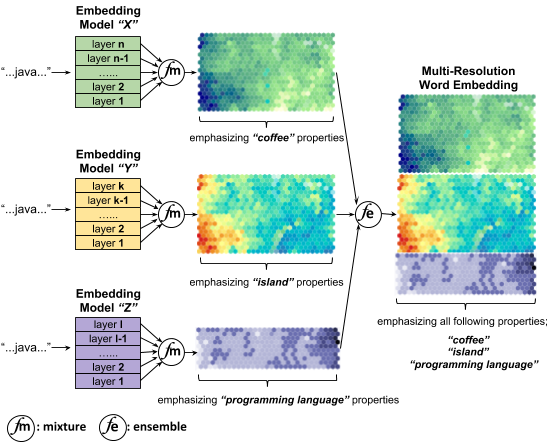}}
\caption{The illustration of multi-resolution word embedding method using an example of \textit{"$\cdots$ java $\cdots$"} }
\label{figure_int}
\end{center}
\vskip -0.2in
\end{figure}

The goal of open domain question answering is to answer questions posed in natural language, using a collection of unstructured natural language documents such as Wikipedia. Given the recent successes of  increasingly sophisticated neural attention based question answering models, \citet{yu2018qanet}, it is natural to break the task of answering a question into two subtasks as suggested in \citet{chen2017reading}:
\begin{itemize}
    \item Retrieval: Retrieval of the document most likely to contain all the information to answer the question correctly.
    \item Extraction: Utilizing one of the above question-answering models to extract the answer to the question from the retrieved document.
\end{itemize}
In our case, we use a collection of unstructured natural language documents as our knowledge base and try to answer the questions without knowing to which documents they correspond. Note that we do not benchmark the quality of the extraction phase; therefore, we do not study extracting the answer from the retrieved document but rather compare the quality of retrieval methods and the feasibility of learning specialized neural models for retrieval purposes. Due to the complexity of natural languages, optimal word embedding, that represents natural language documents in a semantic vector space, is crucial for document retrieval. Traditional word embedding methods learn hierarchical representations of documents where each layer gives a representation that is a high-level abstraction of the representation from a previous layer. Most word embedding methods only use either the  highest layer like Word2Vec by \citet{word2vec_NIPS2013_5021}, or an aggregated representation from the last few layers, such as ELMo by \citet{elmo} as the representation for information retrieval. In this paper, we present a new word embedding approach called \textit{multi-resolution word embedding} that consists of two steps as shown in Figure~\ref{figure_int}. In the first step, we form a mixture of weighted representations across the whole hierarchy of a given word embedding model, so that all resolutions of the hierarchical representation are preserved for the next step. As the second step, we combine all mixture representations from various models as an ensemble representation for the document retrieval task. The proposed word embedding takes advantage of multi-resolution power of individual word embedding models, where each model is trained with a complementary strength due to the diversity of models and corpora. Taking the example of \textit{"$\cdots$ java $\cdots$"} in Figure~\ref{figure_int}, different level of representation of \textit{"java"} including word level (word sense) and concept level (abstract meaning like coffee, island, and programming) are aggregated to form a mixture of representations. In the second step, all these mixture representations from different word embedding models are aggregated to form an ensemble representation, which takes advantage of the complementary strength of individual models and corpora. Consequently, our multi-resolution word embedding delivers the power of multi-resolution with the strength of individual models.  

As another contribution of the paper, we improve the quality of the target document retrieval task by introducing a convolutional residual retrieval network (ConvRR) over the embedding vectors. The proposed ConvRR model further improves the retrieval performance by employing triplet learning with (semi-)hard negative mining on the target corpus.

Our paper is structured as follows: First, we start with a review of recent advances in text embedding in Section \ref{Wordembedding}. In Section \ref{Proposed approach} we describe the details of our approach. More specifically, we describe our multi-resolution word embedding followed by an introduction of a specific deep residual retrieval model that is used to augment text, using the proposed word embedding model for document retrieval. We present an empirical study and compare the proposed method to the baselines that utilize non-augmented word embedding models. In Section \ref{Experiments}, we provide a detailed description of our experiments, including datasets, evaluation metrics, and an implementation. Then, results are reported in Section \ref{Results}. The paper is concluded with some future work in Section \ref{Conclusion}.

\section{Related work}
\label{Wordembedding}
In order to express the importance of a word or a token to a document in a document collection, a numerical statistic is used in information retrieval. The TF-IDF, by \citet{salton1986introduction}, stands for term frequency-inverse document frequency which is proposed to calculate a weighting factor in searches of information retrieval, text mining, and user modeling. Parallel to the advances in the field, new methods that are intended to understand the natural language, are getting proposed. One of the major contributions is called word embedding. There are various types of word embedding in the literature that is well covered by \citet{Sperone}. The influential Word2Vec by \citet{word2vec_NIPS2013_5021} is one of the first popular approaches of word embedding based on neural networks that are built upon the guiding work by \citet{neural_prob} on the neural language model for distributed word representations. This type of implementation is able to conserve semantic relationships between words and their context; or in other terms, surrounding neighboring words. Two different approaches are proposed in Word2Vec to compute word representations. One of the approaches is called Skip-gram that predicts surrounding words, given a target word. The other approach is called Continuous Bag-of-Words that predicts target word, using a bag-of-words context. Global Vectors (GloVe) by \citet{glove},  aims to reduce some limitations of Word2Vec by focusing on the global context instead of surrounding words for learning the representations. The global context is calculated by utilizing the word co-occurrences in a corpus. During this calculation, a count-based approach is functioned, unlike the prediction-based method in Word2Vec. On the other hand, fastText, by \citet{mikolov2018advances}, is also announced recently. It is based on the same principles as others that focus on extracting word embedding from a large corpus. fastText is very similar to Word2Vec except they train high-quality word vector representations by using a combination of known tricks that are, however, rarely used together, which accelerates fastText to learn representations more efficiently. 

The important question still remains on extracting high-quality and more meaningful representations---how to seize the semantic, syntactic and the different meanings in different context---embedding from Language Models (ELMo),by \citet{elmo}, is newly-proposed in order to tackle that question. ELMo extracts representations from a bi-directional Long Short Term Memory (LSTM),by \citet{lstm}, that is trained with a language model (LM) objective on a very large text corpus. ELMo representations are a function of the internal layers of the bi-directional Language Model (biLM) that outputs good and diverse representations about the words/token (a convolutional neural network over characters). ELMo is also incorporating character n-grams, as in fastText, but there are some constitutional differences between ELMo and its predecessors. Likewise, BERT, by \citet{devlin2018bert}, is a method of pre-training language representations that is trained,using a general-purpose "language understanding" model on a large text corpus in an unsupervised manner. Therefore, models, like ELMo and BERT, are contextual uni- or bi-directional models, which generate a representation of each word that is based on the other words in the sentence.

Last, but not least, distance metric learning is designed to amend the representation of the data in a way that retains the related vectors close to each other while separating different ones in the vector space, as stated by \citet{distance_metric_1}, \cite{distance_metric_2}, and \citet{distance_metric_3}. Instead of utilizing a standard distance metric learning, a non-linear embedding of the data, using deep networks, has shown a significant improvement by learning representations using various loss functions, including triplet loss---by \citet{loss_1}, \citet{loss_2}---, contrastive loss---by \citet{loss_3}, \citet{loss_4}---, angular loss---by \citet{loss_5}---, and n-pair loss---by \citet{loss_6}---for influential studies---by \citet{study_use_loss_1}, \citet{study_use_loss_2}, \citet{study_use_loss_3}, and \citet{study_use_loss_4}---.

After providing a brief review of the latest trends in the field, we describe the details of our approach and experimental results in the following sections. 
\section{Proposed approach}\label{Proposed approach}
\subsection{Overview}
\begin{figure}[ht]
\vskip 0.2in
\begin{center}
\centerline{\includegraphics[width=.75\columnwidth]{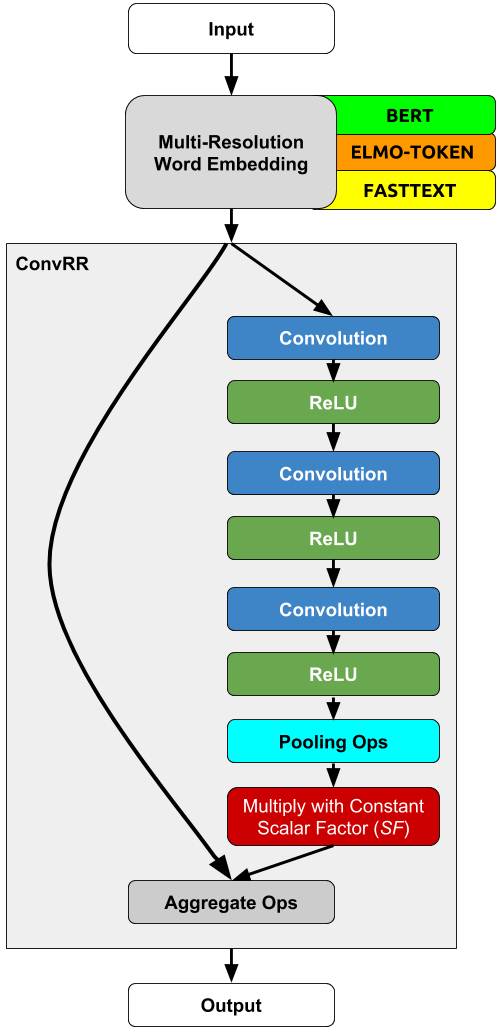}}
\caption{An overview of proposed approach consisting of the multi-resolution word embedding and the \textit{Convolutional Residual Retrieval Network} (\textbf{ConvRR})}
\label{figure_convrr}
\end{center}
\vskip -0.2in
\end{figure}

We describe our proposed approach for document retrieval as follows. First, we devise a new word embedding, called multi-resolution word embedding, which is an ensemble of multi-resolution representations learned from multiple pre-trained word embedding models. Subsequently, a specific neural network model is trained, using a triplet loss. The neural network model is called \textbf{ConvRR}, short for Convolutional Residual Retrieval Network (and alternatively \textbf{FCRR}, short for Full-Connected Retrieval Network by \citet{tolgahan}). The general architecture of the proposed  ConvRR model is shown in Figure~\ref{figure_convrr}. The primary model architecture is not very complex but complex enough to create a semantically more meaningful text embedding on top of a multi-resolution word embedding initialization. The model begins with a series of word inputs $w_1, w_2, w_3, ...., w_k$, that could create a phrase, a sentence, a paragraph and etc. Those inputs, then, are initialized with different resolutions of pre-trained embedding models, including context-free, contextual and numerical analysis. ConvRR further improves the multi-resolution representation by using convolutional blocks through residual connection to the initialized original embedding. The residual connection enables the model not to lose the meaning and the knowledge of the pre-trained multi-resolution embedding but make some adjustments on its knowledge with a limited additional training data. A final representation is then sent to the retrieval task in order to improve the performance.

\subsection{Multi-Resolution Word Embedding}
Since, as aforementioned, existing powerful pre-trained word embeddings are trained using different \textit{data sources} (Wikipedia, Common Craw, and etc.) as well as different \textit{techniques} (supervised, unsupervised or variations). Additionally, pre-trained representations can also be based on context-free either be based on \textit{context-free} (GloVe, FastText, etc.), \textit{contextual} (ELMo, Bert, etc.), and \textit{statistical} (term frequency–inverse document frequency). Contextual representations can further be \textit{unidirectional} or \textit{bidirectional}. Typically, while context-free and statistical word embedding is represented as a vector, contextual word embedding is represented as a matrix. 

Traditionally, one of pre-trained embedding models is selected. A series of word inputs to the network is initialized using the selected embedding model. If the selected embedding model generates a matrix instead of $d$-dimensional vector, then the matrix for each word is represented as follows:
\begin{equation}
\textbf{\emE}_i^j = [\textbf{e}_1, \textbf{e}_2,\cdots, \textbf{e}_l]_{l\times d}
\end{equation}
where $\textbf{E}_i^j\in \mathbb{R}^{l\times d}$ is the $l\times d$-dimensional pre-trained word matrix of $i$-th word input, $j$ denotes the given embedding model and $l$ represents the number of layers in the embedding model. Averaging all the layers (ELMo), $\frac{1}{l}\sum_{i=1}^{l}\textbf{e}_i$, or concatenating each of the last 4 layers (Bert), $<\textbf{e}_{l-3} \oplus \textbf{e}_{l-2} \oplus \textbf{e}_{l-1} \oplus \textbf{e}_l>$ in the matrix are the best practice to create a $d'$-dimensional vector, where $d' = d$ if averaging all layers is used, and $d' = d\times l$ in case concatenating is used.

The proposed multi-resolution word embedding has two cascaded operations: $f_{mixture}(\cdot,\cdot,\cdot)$ and $f_{ensemble}(\cdot)$. 

Forming a mixture of the representations from an embedding model, $f_{mixture}(\cdot,\cdot,\cdot)$, can be formulated as below:
\begin{equation}
\textbf{x}_i^j = f_{mixture}(\textbf{E}_i^j, w_{idf}, \textbf{m}^j)
\end{equation}
where $\textbf{m}^j \in \mathbb{R}^{l}$  is a coefficient vector and $\sum_{i=1}^{l}m^j_i=1$. Each coordinate of the $\textbf{m}^j$ represents a magnitude to weight the corresponding layer of the model $\textbf{E}_i^j$. $w_{idf}$ denotes an IDF weight of the $i$-th word input. $f_{mixture}(\cdot,\cdot,\cdot)$ is an aggregate function, which aggregates the input using an operation such as $sum$, $average$, and $concatenate$. Weighted layers of the model $\textbf{E}_i^j$ are then computed by that aggregate function. $x_i^j$ is the $d'$-dimensional vector where $d'=d$ if $f_{mixture}(\cdot,\cdot,\cdot)$ is defined by $sum$ or $average$, and $d' = d\times l$ in case $f_{mixture}(\cdot,\cdot,\cdot)$ is defined by $concatenate$. The obtained mixture of representations from multiple word embedding models can form an ensemble representation as follows. 
\begin{equation}
X'_i = \{\textbf{x}_i^1, \textbf{x}_i^2, \cdots \textbf{x}_i^n\}
\end{equation}
where $X'_i$ is a set of representations from different embedding models, using $f_{mixture}(\cdot,\cdot,\cdot)$ for the $i$-th word input and $n$ is the number of embedding models. $f_{ensemble}(\cdot)$ is a function to aggregate all representations in $X'_i$ and can be defined as follows:
\begin{equation}
\textbf{x}_i = f_{ensemble}(X'_i, \textbf{u})
\end{equation}
% \begin{gather*} 
% \text{or, in other terms;} \\
% x_i = f(E_i^1, w_{idf}, m_1, \gamma_1) \Delta \cdots \Delta f(E_i^j, w_{idf}, m_j, \gamma_j)
% \end{gather*}
where $f_{ensemble}(\cdot)$ is also a aggregate function defined by an operation like $sum$, $average$, and $concatenate$. Note that, representations are coerced to a common length, if $f_{ensemble}(\cdot)$ is defined by $sum$ or $average$. Additionally, $\textbf{u} \in \mathbb{R}^{n}$ is a coefficient vector and $\sum_{j=1}^{n}\frac{u^j}{|| u||}=1$. Each coordinate of the $\textbf{u}$ represents a magnitude to weight the corresponding embedding model of the multi-resolution word embedding model. Hence, $\textbf{x}_i$ is $d''$-dimensional multi-resolution word embedding of the $i$-th word input. The pseudo-code of the proposed approach is shown in Algorithm~\ref{alg:multi-resolution}.

\begin{algorithm}[ht]
   \caption{Multi-Resolution Word Embedding for the $i$-th word input}
   \label{alg:multi-resolution}
\begin{algorithmic}
   \STATE {\bfseries Input:}  \\ \text{idf weight:} $w_{idf}$ \\ \text{set of embedding models:} $\textbf{\emE}_i$ = $\{\textbf{\emE}_i^1, \textbf{\emE}_i^2, \cdots, \textbf{\emE}_i^n\}$, \\ \text{set of coefficient vectors, one for each model:} \\$\textbf{\emM}$ = $\{\textbf{m}^1, \textbf{m}^2, \cdots, \textbf{m}^n\}$, \\ \text{vector of coefficient values, one for each model:}
   \\$\textbf{u}$ = $\ \left[u^1, u^2, \cdots, u^n\ \right]$
   \STATE {\bfseries Output:} \\ \text{multi-resolution word embedding:} $\textbf{x}_i$
   \STATE {\bfseries Begin}
   \STATE $X'_i$ = \{\}
   \FOR{$j=1$ {\bfseries to} $n$}
   \STATE $\textbf{x}_i^j$ = $f_{mixture}($$\textbf{E}_i^j$, $w_{idf}$, $\textbf{m}^j$$)$
   \STATE $\textbf{x}_i^j$ is added to $X'_i$
   \ENDFOR
   \STATE $\textbf{x}_i$ = $f_{ensemble}($$X'_i$$, \textbf{u})$
   \STATE {\bfseries Return:} $\textbf{x}_i$
   \STATE {\bfseries End}
\end{algorithmic}
\end{algorithm}

With the multi-resolution word embedding approach, we are generating embedding by taking the following aspect into consideration: 
\begin{itemize}
    \item \underline{Multi-sources}: Instead of relying on one pre-trained embedding model, we want to utilize the power of multiple pre-trained embedding models since they are trained using different data source as well as different techniques. Therefore, integrating different word embedding models can harness the complementary power of individual models.   
    \item \underline{Different layers}: We take the embedding from different layers of $\textbf{\emE}$ each embedding model instead of just the last layer or few top layers.
    \item \underline{Weighted embedding}: Incorporating word embedding with an inverse document frequency (\textit{IDF}) produce better results for information retrieval and text classification as presented by \citet{deboom}. An IDF is formulated as:  $log_e(\frac{\# of documents}{df_w})$, where 
    %a term frequency (\textit{tf}) is the number of times the word occurs in a particular document, and 
    a document frequency ($df_w$) is the number of documents in the considered corpus that contain that particular word $w$. 
\end{itemize}

\subsection{ConvRR}
To further improve the performance of document retrieval, a convolutional residual retrieval (ConvRR) model is trained on top of the proposed multi-resolution word embedding. The model is presented in Figure~\ref{figure_convrr}. Let $\textbf{x}_i \in \mathbb{R}^{d''}$ be the $d''$-dimensional proposed multi-resolution word embedding of the $i$-th word input in a text; therefore, the word inputs can be denoted as a matrix: 
\begin{equation}
\textbf{X} = [\textbf{x}_1, \textbf{x}_2, \textbf{x}_3, \cdots, \textbf{x}_k]_{k\times d''}
\end{equation}
where $k$ is the number of word inputs in a text. The ConvRR generates feature representations, which can be expressed as the following: 
\begin{equation}
\textbf{X}'' = f(\textbf{W}, \textbf{X}, sf)
\end{equation}
\begin{equation}
\textbf{o}  = \textbf{X}'' +  \frac{1}{k}\sum_{i=1}^{k}\textbf{x}_i 
\end{equation}
where $f(\cdot,\cdot,\cdot)$ is the convolutional residual retrieval network that executes series of convolutional components (a convolution and a rectified linear unit (ReLU) \citet{Nair:2010}), a pooling, and a scaling operation. $\textbf{X}''$ is produced by multi-resolution word embedding X with trainable weights $\textbf{W} \in \mathbb{R}^{d'' \times ws \times d''}$. The weight matrix $\textbf{W}$ contains $d''$ kernels, each of them has $ws \times d''$, convolving $\textbf{ws}$ contiguous vectors. $ws$ and $d''$ represent window-size and number of kernels respectively. Average pooling operation is added after final convolutional component, which can consolidate some unnecessary features and boost computational efficiency. $sf$ is a scaling factor that weights the output with a constant factor. Hence, $\textbf{X}''$ is trained on how much contribution it adds to the $\textbf{X}$ using residual connection to improve the retrieval task.  Final output $\textbf{o} = [o_1,o_2, o_3, \cdots, o_{d''}] \in\mathbb{R}^{d''}$ is generated, which will be fed into the next component. Note that each of feature vector $\textbf{o}$ is normalized to unit $l_2$ norm before passing to the next step.

\begin{figure}[ht]
\vskip 0.2in
\begin{center}
\centerline{\includegraphics[width=.80\columnwidth]{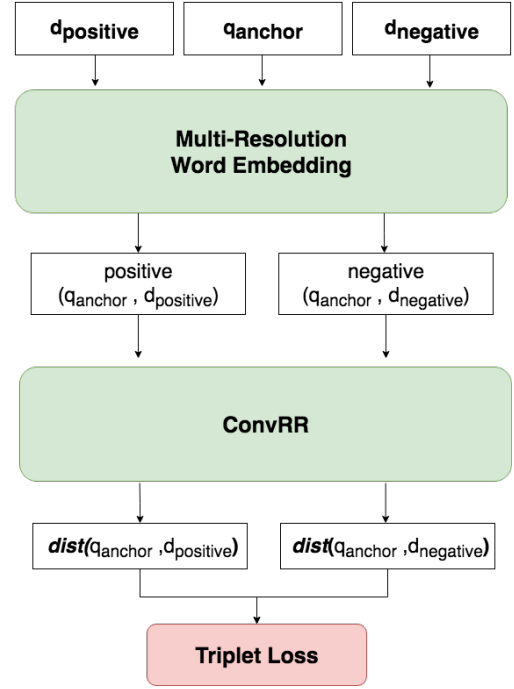}}
\caption{Overall flow diagram for the proposed approach}
\label{figure_overall}
\end{center}
\vskip -0.2in
\end{figure}

\subsection{Loss Function}
In order to train the ConvRR network to perform well on retrieval task and generalize well on unseen data, we utilize the Siamese architecture with \textit{triplet loss} during the training period as shown in Figure~\ref{figure_overall}. With this setup, the network is encouraged to reduce distances between positive pairs so that they are smaller than negative ones. A particular question $\vq_\text{anchor}$ would be a question close in proximity to a document $\vd_\text{positive}$ as the positive pair to the same question than to any document $\vd_\text{negative}$ as they are positive pairs to other questions. The key point of the $\mathscr{L}_\text{triplet}$ is to build the correct triplet structure which should meet the condition of the following equation:

$||\vq_\text{anchor},\vd_\text{positive}|| + m < ||\vq_\text{anchor}, \vd_\text{negative}||$ 

For each anchor, the positive $\vd_\text{positive}$ is selected in such a way $\argmax_{\vd_\text{positive}}||\vq_\text{anchor},\vd_\text{positive}||$ and likewise the hardest negative $\vd_\text{negative}$ in such a way that $\argmin_{\vd_\text{negative}}||\vq_\text{anchor},\vd_\text{negative}||$ to form a triplet. This triplet selection strategy is called \textit{hard triplets mining}. 

Let $\emT = (\vd_\text{positive}, \vq_\text{anchor}, \vd_\text{negative})$ be a triplet input. Given $\emT$, the proposed approach computes the distances between the positive and negative pairs via a two-branch siamese subnet through the multi-resolution word embedding and ConvRR.
\begin{equation}
\mathscr{L}_\text{triplet} = [||\vq_\text{anchor},\vd_\text{positive}|| - ||\vq_\text{anchor},\vd_\text{negative}|| + m]^+
\end{equation}
where $m > 0$ is a scalar value, namely margin.

\section{Experiments}\label{Experiments}
\subsection{Datasets}
In order to evaluate our proposed approach, we conducted extensive experiments on two large question-answering datasets, including SQuAD \citet{squad}, and QUASAR \citet{dhingra2017quasar}. 
\subsubsection{SQuAD}
The Stanford Question Answering Dataset (SQuAD) \citet{squad} is a large reading comprehension dataset that is built with $100,000+$ questions. Each of these questions are composed by crowdworkers on a set of Wikipedia documents, where the answer to each question is a segment of text from the corresponding reading passage. In other words, the consolidation of retrieval and extraction tasks are aimed at measuring the success of the proposed systems.

\subsubsection{QUASAR}
The Question Answering by Search And Reading (QUASAR) is a large-scale dataset consisting of QUASAR-S and QUASAR-T. Each of these datasets are built to focus on evaluating systems devised to understand a natural language query, large corpus of text and to extract answer to the question from that corpus. Similar to SQuAD, the consolidation of retrieval and extraction tasks are aimed at measuring the success of the proposed systems. Specifically, QUASAR-S comprises $37,012$ fill-in-the-gaps questions that are collected from the popular website Stack Overflow, using entity tags. Since our research is not about addressing fill-in-the-gaps questions, we want to pay attention to the QUASAR-T dataset that fulfill the requirements of our focused retrieval task. The QUASAR-T dataset contains $43,012$ open-domain %trivia
questions collected from various internet sources. The candidate documents for each question in this dataset are retrieved from an Apache Lucene based search engine built on the ClueWeb09 dataset \citet{callan2009clueweb09}.

The number of queries in each dataset, including their subsets, is listed in Table~\ref{dataset_statistic}. 
\begin{table}[ht]
\caption{Datasets Statistics: Number of queries in each train, validation, and test subsets}
\label{dataset_statistic}
\vskip 0.001in
\begin{center}
\begin{small}
\begin{sc}
\begin{tabular}{lccccr}
\toprule
Dataset & train & valid. & test & total \\
\midrule
SQuAD & 87,599 & 10,570 & hidden& 98,169+\\
QUASAR-T & 37,012& 3,000& 3,000& 43,012\\
\bottomrule
\end{tabular}
\end{sc}
\end{small}
\end{center}
\vskip -0.1in
\end{table}

\subsection{Evaluation}
The retrieval model aims to improve the $recall@k$ score by selecting the correct pair among all candidates. Basically, $recall@k$ would be defined as the number of correct documents as listed within top-$k$ order out of all possible documents, \citet{Manning:2008:IIR:1394399}. Additionally, embedding representations are visualized, using t-distributed stochastic neighbor embedding by \citet{Maaten2008VisualizingDU} in order to project the clustered distributions of the questions that are assigned to same documents.
\subsection{Implementation}
\subsubsection{Input}
Word embedding were adopted, using the proposed multi-resolution word embedding. $f_{mixture}(\cdot,\cdot,\cdot)$ and $f_{ensemble}(\cdot)$ settings that represent the best configuration are shown in Table~\ref{multi_resolution_mixture} and Table~\ref{multi_resolution_ensemble} respectively.

\begin{table}[ht]
\caption{$f_{mixture}(\cdot,\cdot,\cdot)$ configuration of the multi-resolution word embedding}
\label{multi_resolution_mixture}
\vskip 0.001in
\begin{center}
\begin{small}
\begin{sc}
\begin{tabular}{lcccr}
\toprule
%function name & \textbf{E} & w_{idf} & \textbf{m} & \gamma \\
\textbf{E} & $w_{idf}$ & \textbf{$m$} & $f_{mix}$ & out \\
\midrule
Bert & False & [$\frac{1}{4}$, $\frac{1}{4}$, $\frac{1}{4}$, $\frac{1}{4}$,0,..,0] & $concat.$ &$\textbf{x}^1$\\
ELMo & True & [0, 0, 1] & $sum$ &$\textbf{x}^2$\\
FastText & True & [1] & $sum$  &$\textbf{x}^3$\\
\bottomrule
\end{tabular}
\end{sc}
\end{small}
\end{center}
\vskip -0.1in
\end{table}

\begin{table}[ht]
\caption{$f_{ensemble}(\cdot)$ configuration of the multi-resolution word embedding}
\label{multi_resolution_ensemble}
\vskip 0.001in
\begin{center}
\begin{small}
\begin{sc}
\begin{tabular}{lcr}
\toprule
\textbf{X'} & \textbf{$u$} & $f_{ensemble}$ \\
\midrule
\{$\textbf{x}^1$, $\textbf{x}^2$, $\textbf{x}^3$\}& [$\frac{1}{3}$, $\frac{1}{3}$, $\frac{1}{3}$] & $concat.$\\
\bottomrule
\end{tabular}
\end{sc}
\end{small}
\end{center}
\vskip -0.1in
\end{table}

The short form of this multi-resolution word embedding is called as follows:  \textit{BERT $\oplus$ ETwI $\oplus$ FTwI} where $(.)wI$ is denoting \textit{"with IDF"} and $\oplus$ represents concatenation operation. The dimension of embedding is $4,372$. 

\subsubsection{ConvRR Cofiguration}
ConvRR is trained, using ADAM optimizer, by \citet{adam}, with a learning rate of $10^{-3}$. For the sake of equal comparison, we fixed the seed of randomization. We also observed that $10^{-3}$ as a weight decay is the reasonable value to tackle over-fitting. We choose windows-size $ws=5$, number of kernel $d''= 4,372$, and the scaling factor $sf = 0.05$. We trained the network with $400$ iterations with a batch size of $2,000$ using a triplet loss with a margin $m=1$. Note that the best performance is achieved using a relative large batch size. All experiments are implemented with Tensorflow 1.8+ by \citet{tensorflow2015-whitepaper} on 2 $\times$ NVIDIA Tesla K80 GPUs.

\section{Results}\label{Results}
First, we study different embedding models. We initialize text inputs of datasets, using different traditional embedding models. Additionally, we also initialize text inputs, using the proposed multi-resolution word embedding. We configured the multi-resolution word embedding for different embedding models. We compared our model with the following baselines: TF-IDF, BERT, ELMo-LSTM1 (first layer), ELMo-ELMO (averaging all layers), ELMo-LSTM2 (second layer), GloVe, ELMo-TOKEN (token layer), fastText, BERT with IDF weight, ELMo-TOKEN with IDF weight, fastText with IDF weight, multi-resolution word embedding with a concatenation of ELMo-TOKEN layer with IDF weight and fastText with IDF weight and finally a concatenation of BERT \textit{(concatenation of last 4 layer representations)}, ELMo-TOKEN layer with IDF weight, and  fastText with IDF weight.

The $recall@k$ results that calculated for SQuAD and QUASAR-T datasets are listed in Table~\ref{table-squad} and Table~\ref{table-quasar}. Our ConvRR model initialized with the proposed multi-resolution word embedding outperforms all the baseline models on these datasets. 

\begin{table}[ht]
\caption{Experimental results on SQUAD. $recall@k$ retrieved documents, using different models and the proposed approach.}
\label{table-squad}
\vskip 0.15in
\begin{center}
\begin{small}
\begin{sc}
\begin{tabular}{lccccr}
\toprule
Embedding/Model & @1 & @3 & @5 \\
\midrule
\multicolumn{3}{l}{\small \textbf{Base Embeddings}} \\ 
\midrule
TF-IDF & 8.77& 15.46& 19.47\\
BERT & 18.89& 32.31& 39.52\\
ELMO-AVG & 21.24& 36.24& 43.88\\
GLOVE & 30.84& 47.14& 54.01\\
FASTTEXT & 42.23& 59.86& 67.12\\
\midrule
\multicolumn{3}{l}{\small \textbf{Multi-Resolution Emb. (w/o  Ensemble)}} \\ 
\midrule
ELMO-LSTM1 & 19.65& 34.34& 42.52\\
BERT w/ IDF & 21.81& 36.35& 43.56\\
ELMO-LSTM2 & 23.68& 39.39& 47.23\\
ELMO-TOKEN & 41.62& 57.79& 64.36\\
ELMO-TOKEN w/ IDF & 44.85& 61.55& 68.07\\
FASTTEXT w/ IDF & 45.13& 62.80& 69.85\\
\midrule
\multicolumn{3}{l}{\small \textbf{Multi-Resolution Emb. (w/ Ensemble)}} \\ 
\midrule
ETwI $\oplus$ FTwI & 46.33& 63.13& 69.70\\
BERT $\oplus$ ETwI $\oplus$ FTwI & 48.49& 64.96& 71.05\\
\midrule
\multicolumn{3}{l}{\small \textbf{Base Embedding + Downstream Models}} \\ 
\midrule
FASTTEXT + FCRR & 45.7& 63.15& 70.02\\
FASTTEXT + ConvRR & 47.14& 64.16& 70.87\\
\midrule
\multicolumn{3}{l}{\small \textbf{Multi-Resolution Emb. + Downs. Models}} \\ 
\midrule
\tiny BERT $\oplus$ ETwI $\oplus$ FTwI + FCRR & 50.64& 66.16& 73.44\\
\tiny BERT $\oplus$ ETwI $\oplus$ FTwI + ConvRR & \textbf{52.32}& \textbf{68.26}& \textbf{75.68}\\
\bottomrule
\end{tabular}
\end{sc}
\end{small}
\end{center}
\vskip -0.1in
\end{table}

\begin{table}[ht]
\caption{Experimental results on QUASAR-T. $recall@k$ retrieved documents, using different models and the proposed approach.}
\label{table-quasar}
\vskip 0.15in
\begin{center}
\begin{small}
\begin{sc}
\begin{tabular}{lccccr}
\toprule
Embedding/Model & @1 & @3 & @5 \\
\midrule
\multicolumn{3}{l}{\small \textbf{Base Embeddings}} \\ 
\midrule
TF-IDF & 13.86 & 20.2& 23.13\\
BERT & 25.5& 34.2& 37.86\\
ELMO-AVG & 27.93& 37.86& 42.33\\
GLOVE & 32.63& 40.73& 44.03\\
FASTTEXT & 46.13& 56.00& 59.46\\
\midrule
\multicolumn{3}{l}{\small \textbf{Multi-Resolution Emb. (w/o  Ensemble)}} \\ 
\midrule
ELMO-LSTM1 & 24.6& 33.01& 36.9\\
ELMO-LSTM2 & 27.03& 36.33& 40.56\\
BERT w/ IDF & 27.33& 38.43& 40.11\\
ELMO-TOKEN & 44.46& 54.86& 59.36\\
ELMO-TOKEN w/ IDF & 48.86& 60.56& 65.03\\
FASTTEXT w/ IDF & 49.66& 58.70& 61.96\\
\midrule
\multicolumn{3}{l}{\small \textbf{Multi-Resolution Emb. (w/ Ensemble)}} \\ 
\midrule
ETwI $\oplus$ FTwI & 48.78& 60.05& 64.10\\
BERT $\oplus$ ETwI $\oplus$ FTwI & 49.46& 60.93& 65.66\\
\midrule
\multicolumn{3}{l}{\small \textbf{Base Embedding + Downstream Models}} \\ 
\midrule
FASTTEXT + FCRR & 47.11& 58.25& 62.12\\
FASTTEXT + ConvRR & 48.17& 59.06& 63.07\\
\midrule
\multicolumn{3}{l}{\small \textbf{Multi-Resolution Emb. + Downs. Models}} \\ 
\midrule
\tiny BERT $\oplus$ ETwI $\oplus$ FTwI + FCRR & 49.55& 61.58& 64.53\\
\tiny BERT $\oplus$ ETwI $\oplus$ FTwI + ConvRR & \textbf{50.67}& \textbf{63.09}& \textbf{67.38}\\
\bottomrule
\end{tabular}
\end{sc}
\end{small}
\end{center}
\vskip -0.1in
\end{table}

\begin{figure}[ht]
\vskip 0.2in
\begin{center}
\centerline{\includegraphics[width=.9999\columnwidth]{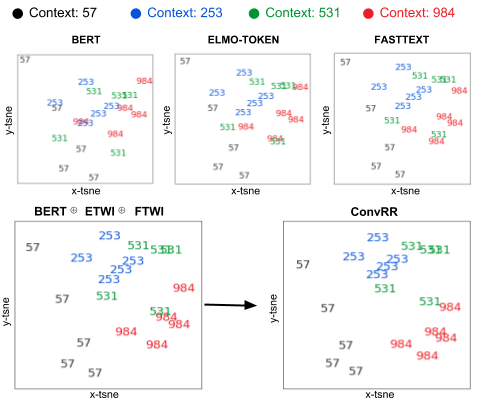}}
\caption{t-SNE map visualizations of various embedding models for all question representations of 4 (57, 253, 531, 984) sampled contexts/documents that are extracted from SQuAD validation dataset. }
\label{figure_tsne}
\end{center}
\vskip -0.2in
\end{figure}

The t-SNE visualization of question embeddings that are derived, using different embedding models, including BERT, ELMo-TOKEN layer, fastText, multi-resolution word embedding with  a concatenation of BERT \textit{(concatenation of last 4 layer representations)}, ELMo-TOKEN layer with IDF weight, and  fastText with IDF weight, and ConvRR are shown in Figure~\ref{figure_tsne}. Note that those questions match the particular 4 (labeled as 57, 253, 531, 984) sampled contexts/documents that are extracted from SQuAD validation dataset. The visualization shows that the proposed multi-resolution word embedding significantly improves the clustering of the questions and corresponding contexts/documents. The result is further improved by using ConvRR, the proposed retrieval model. 

\section{Conclusion}\label{Conclusion}
We developed a new multi-resolution word embedding approach, which harnesses the power of individual strength of diverse word embedding methods. The performance of the proposed approach is further improved by using a convolutional residual retrieval model optimized using a triplet loss function for the task of document retrieval, which is a crucial step for many Natural Language Processing and information retrieval tasks. We further evaluate the proposed method for document retrieval from an unstructured knowledge base. The empirical study using large datasets including SQuAD and QUASAR benchmark datasets shows a significant performance gain in terms of the recall. In the future, we plan to apply the proposed framework for other information retrieval and ranking tasks. We also want to improve the performance of the retrieval task by applying and developing new loss functions and retrieval models.

\nocite{langley00}

\bibliography{example_paper}
\bibliographystyle{icml2019}

\end{document}